\pdfoutput=1
\documentclass[%
 aip,
 amsmath,amssymb,
 reprint,%
]{revtex4-1}

\usepackage{graphicx}
\usepackage{bm}

\usepackage[utf8]{inputenc}
\usepackage[T1]{fontenc}
\usepackage{mathptmx}
\usepackage{etoolbox}

\makeatletter
\def\@email#1#2{%
 \endgroup
 \patchcmd{\titleblock@produce}
  {\frontmatter@RRAPformat}
  {\frontmatter@RRAPformat{\produce@RRAP{*#1\href{mailto:#2}{#2}}}\frontmatter@RRAPformat}
  {}{}
}%
\makeatother
\begin{document}

\preprint{AIP/123-QED}

\title[Effect of a femtosecond-scale temporal structure of a laser driver on generation of betatron radiation]{Effect of a femtosecond-scale temporal structure of a laser driver on generation of betatron radiation by wakefield accelerated electrons}
\author{A. D. Sladkov}
\author{A. V. Korzhimanov}%
 \email{artem.korzhimanov@ipfran.ru}
 \altaffiliation[Also at ]{Lobachevsky State University of Nizhny Novgorod, Russia.}
\affiliation{ 
Institute of Applied Physics of the Russian Academy of Sciences, Nizhny Novgorod, Russia
}

\date{\today}

\begin{abstract}
A brightness of a betatron radiation generated by laser wakefield accelerated electrons can be increased by utilizing the laser driver with shorter duration at the same energy. Such shortening is possible by pulse compression after its nonlinear self-phase modulation in thin plate. However, this method can lead to a rather complex femtosecond-scale time structure of the pulse. In this work, the results of numerical simulations show that the presence of a pedestal or prepulses containing few percents or more of the main pulse energy can inhibit the acceleration process and lead to lower energies of accelerated electrons and generated photons. Simultaneously, the presence of a relatively long and intense postpulse following the main pulse at optimal distance can enhance the oscillations of accelerating electrons and increase the brightness of betatron radiation.
\end{abstract}

\maketitle

\section{\label{sec:intro}Introduction}

Soft X-ray sources are widely used in many applications \cite{Seeck_Murphy_2015, sanchez2021x}. One of the urgent modern problems is the creation of sufficiently bright sources of soft X-ray radiation for applications such as absorption X-ray radiography and X-ray phase-contrast imaging for biological and medical purposes \cite{Banerjee2015, Flegentov2021, Davis1995, Najmudin2014}. As a rule, to generate bright radiation in this range, radiofrequency accelerators such as synchrotrons or linear accelerators, followed by an undulator, are used. However, to generate bright radiation in the range of 1--100 keV, electron beams with an energy of hundreds of MeV are required, which requires the large size of conventional accelerators. Recently, much more compact laser-plasma X-ray sources has been proposed as an alternative for those applications \cite{Corde2013,Albert2014,Umstadter2015}. Among them, the most popular method of radiation generation in the required range is the so-called betatron source \cite{Kostyukov2003,Kiselev2004}. In this method, the X-rays are radiated by electrons simultaneously being accelerated and transversely oscillating in the plasma cavity behind an intense laser pulse propagating in underdense plasmas.

Despite the fact that the brightness of the betatron source was experimentally achieved at the level of $10^{8}$--$10^{9}$ photons per shot, the problem of further increasing the brightness is still relevant \cite{Albert2016}. One of the possible ways to increase the brightness for the same laser pulse energy can be to reduce the duration of the driving laser pulse.

Indeed, according to similarity theory \cite{Gordienko2005}, a waist of the laser pulse $w_0$ should be consistent with its intensity and a density of the plasma target due to a relation $(kw_0)^2 \approx 2a_0/n_0$, where $k$ is a wavenumber of the laser pulse, $a_0 = eE_0/m\omega c$ is a relativistic amplitude ($e$ -- elementary charge, $E_0$ -- electric field amplitude of the laser pulse, $m$ -- electron mass, $\omega$ -- laser frequency, $c$ -- speed of light), $n_0 = e^2 N_e/\varepsilon_0m\omega^2$ is a plasma overdense parameter ($N_e$ -- background electron concentration, $\varepsilon_0$ -- electrical constant). In this case, the energy gained by an electron in a plasma wake is scaled as $\gamma \approx 2a_0/3n_0$, where $\gamma$ is the maximum achievable relativistic gamma-factor of the accelerated electrons. Let us fix the laser pulse energy: $W = W_{\rm rel}a_0^2(kw_0)^2\omega\tau_0$, where $W_{\rm rel} = \varepsilon_0m^2c^5/2e^2\omega$ ($W_{\rm rel}[\mathrm{J}] = 1.84\times 10^{-7}\lambda[\mathrm{\mu m}]$), $\tau_0$ is the pulse duration, and find the dependence of the electron energy on the pulse duration and plasma density:
\begin{equation}\label{gamma}
    \gamma \approx \frac{1}{3}\left(\frac{4W}{W_{\rm rel}n_0^2\omega\tau_0}\right)^{1/3}
\end{equation}

The power of the betatron radiation can be estimated from the following expression \cite{Kostyukov2003}:
\begin{equation}\label{Pb}
    P_b \approx \frac{e^2\omega^4}{12c^3}\gamma^2 n_0^2 r_b^2
\end{equation}
where $r_b$ is the amplitude of betatron oscillations, which we will consider in accordance with the theory of ultrarelativistic similarity \cite{Gordienko2005} as inversely proportional to the root of the plasma density: $r_b = n_0^{-1/2}r_{b0}$, where $r_{b0}$ is a constant determined by the shape of the laser pulse. Then from (\ref{gamma}) and (\ref{Pb}) we get:
\begin{equation}
    P_b \approx \frac{e^2\omega^4r_{b0}^2}{108c^3}\left(\frac{4W}{W_{\rm rel}}\right)^{2/3}\left(\frac{n_0}{(\omega\tau_0)^2}\right)^{1/3}.
\end{equation}
From this it can be seen that at the fixed energy and the duration of the laser pulse, the power of betatron radiation increases with increasing plasma density. However, it is impossible to increase the plasma density indefinitely because as the plasma density increases, the plasma wavelength decreases, and when this length becomes shorter than the pulse duration, the acceleration efficiency drops sharply, so the most optimal is to use plasma with a density of $n_0 \approx 1/(\omega\tau_0)$. In this case, we have the betatron radiation power equal to
\begin{equation}
    P_b^{\rm opt} \approx \frac{e^2\omega^3r_{b0}^2}{108c^3\tau_0}\left(\frac{4W}{W_{\rm rel}}\right)^{2/3}
\end{equation}
Thus, at fixed pulse energy and an optimal choice of the plasma density and the laser pulse waist, the betatron radiation power increases with decreasing laser pulse duration.

Since it is possible to increase the brightness of betatron sources at the same energy of the laser pulse by shortening it, it seems promising to use the recently implemented scheme for compressing multiterawatt and petawatt laser pulses using nonlinear self-phase modulation in thin plates \cite{Khazanov2019,Ginzburg2019,Ginzburg2020QE,Ginzburg2020PRA,Mironov2020,Shaykin2021,Kim2022}. In this case, the pulse energy almost conserves, whilst the duration can be reduced several times. Under the best conditions, pulses with a duration of 11 fs at a power of 1.5 PW were achieved in this way \cite{Ginzburg2021}. The use of the pulses compressed by this method was also recently demonstrated experimentally to increase the brightness of betatron radiation \cite{Fourmaux2022}. However, as modeling shows, such pulses can have a rather complex time structure at femtosecond scales. In particular, a pulse can have a femtosecond pedestal and be accompanied by pre- and postpulses containing up to 20--30~\% of the pulse energy. Thus, there is a need in studying what effect such a structure of the laser pulse can have on the process of electron acceleration in a plasma wake, as well as on the generation of betatron radiation. This work is devoted to the analysis of this problem.

\section{\label{sec:numerics}Parameters of numerical simulations}

The analysis was carried out by numerical simulations by the fully electrodynamic Particle-In-Cell method using the PICADOR code \cite{Surmin2016}. The simulations were carried out in 2D geometry in a box $120\times120$ $\mu$m$^2$ in size. The step size in the longitudinal direction was $\Delta x=0.05$ $\mu$m, and in the transverse direction it was $\Delta y=1$ $\mu$m. The time step was $\Delta t = \Delta x/2c = 8.34\times10^{-2}$ fs.

A laser pulse with a wavelength of 900 nm and polarized in out-of-plane $z$-direction was generated at the left boundary. The pulse had a Gaussian transverse profile with a radius of 15 $\mu$ m at the level $1/e$, which corresponds to a focusing spot waist of 17.6 $\mu$m at FWHM (Full Width at Half Maximum) of the intensity. The longitudinal profile of the pulse varied.

The plasma was assumed to be homogeneous with an electron concentration equal to $4.13\times10^{18}$ cm$^{-3}$, which corresponds to the overdense parameter $n_0=0.003$. The plasma boundary was initially located at a distance of 10 $\mu$m from the left boundary of the box.

To reduce the calculation time, the moving window technique was used. Window movement began at the time instant $t=290$ fs after the beginning of the calculation, which corresponded to the position of the laser pulse front at a distance of 75–90 $\mu$m from the left boundary. The speed of window movement was equal to the speed of light. The total calculation time was 10 ps.

Betatron radiation was calculated using the Monte Carlo method using the expressions for synchrotron radiation, as described in the paper \cite{Gonoskov2015}.

\section{\label{sec:results}Results}

\subsection{Comparison of long and short laser pulses}
First, we carried out simulations confirming that the use of a shorter laser pulse increases the brightness of the betatron source. To do this, we compared the case of a 50-fs pulse and an 11-fs pulse. Both pulses had a Gaussian longitudinal profile and the same energy equal to 3.9 J, which roughly corresponds to the parameters obtained experimentally \cite{Fourmaux2022}. Here and below, the duration was determined at FWHM of the intensity. Thus, the dimensionless pulse amplitudes were $a_0 = 3.5$ and 7.5, respectively.

\begin{figure*}
\centering
\includegraphics[width=1\textwidth]{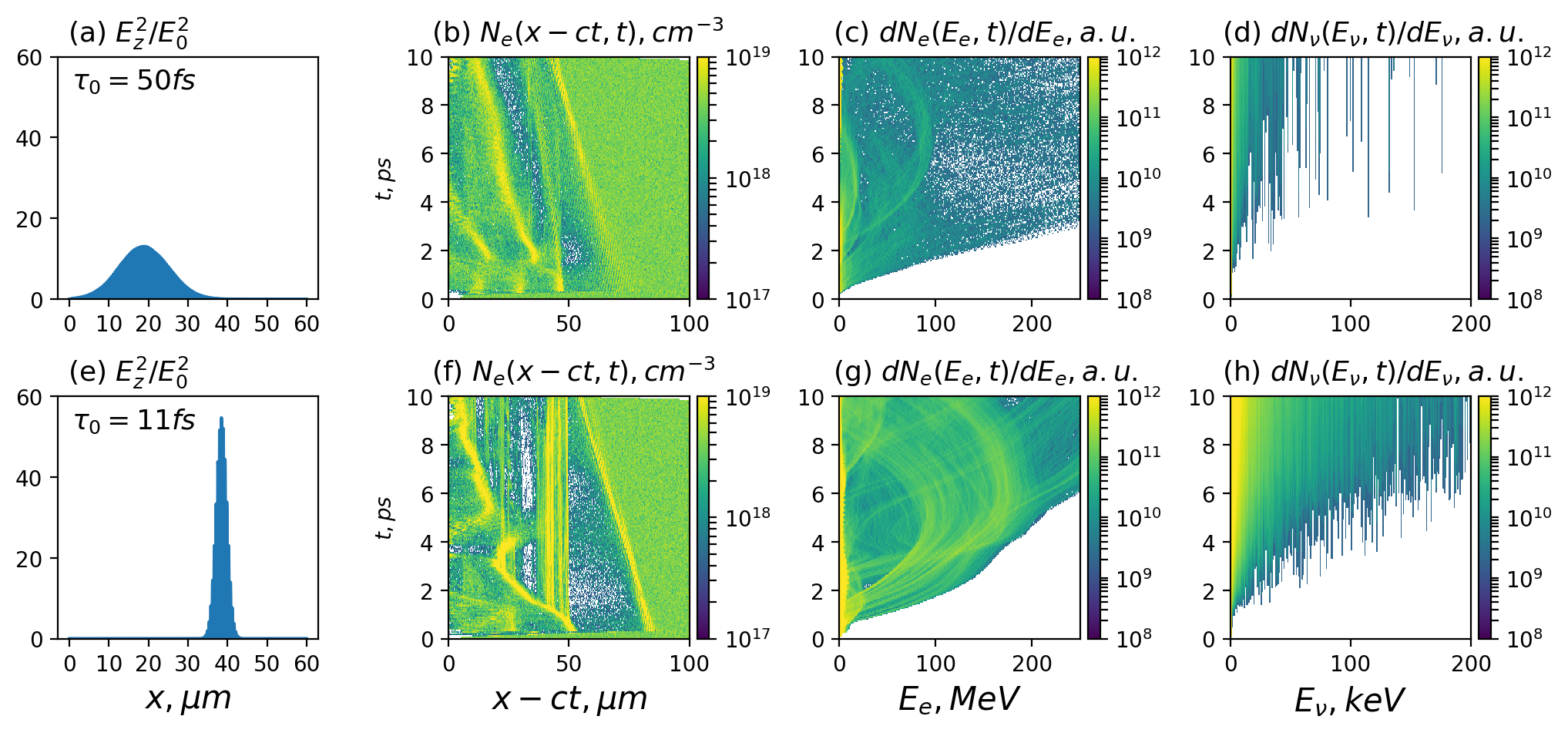}
\caption{\label{fig:50vs11}Comparison of simulation results for Gaussian laser pulses with a duration of 50 fs and 11 fs. (a), (e) --- time profile of the incident pulses; (b), (f) --- spatio-temporal dynamics of the electron concentration on the central axis of the system in the moving window; (c),(g) --- time dependence of the electron energy spectrum in the moving window; (d), (h) --- time dependence of the photon energy spectrum in the moving window.}
\end{figure*}

The simulation results are shown in Fig.~\ref{fig:50vs11}. It can be seen that in both cases a plasma wake with strong cavitation is formed behind the laser pulse, into which the electron beam is trapped. This beam is accelerated to an energy of more than 100 MeV and, oscillating in plasma fields, generates radiation in the range up to 200 keV. For longer pulses, however, the acceleration process is noticeably less efficient, and the generated spectra are much narrower.

\subsection{Influence of a pedestal}
Next, we studied the influence of a femtosecond pedestal on the interaction process. In this series of calculations, the main pulse had a Gaussian shape and a duration of 11 fs. The pedestal also had a Gaussian shape, and its duration was 50 fs. The centers of the main pulse and the pedestal coincided. The amplitude of the pedestal was varied in the range from $a_0=0.4$ to $a_0=3.3$, while the amplitude of the main pulse was also changed so that the total amplitude of the pulse remained constant and equal to $a_0 = 7.5$. The total energy of the pulse and the pedestal varied in the range from 3.2 J at the pedestal amplitude $a_0=1.7$ to 4.6 J at the pedestal amplitude $a_0=3.3$. Thus, the energy variation was relatively small, whereas the constancy of the pulse amplitude made it possible to expect the constancy of the energy of the accelerated electrons.

\begin{figure}
\centering
\includegraphics[width=0.5\textwidth]{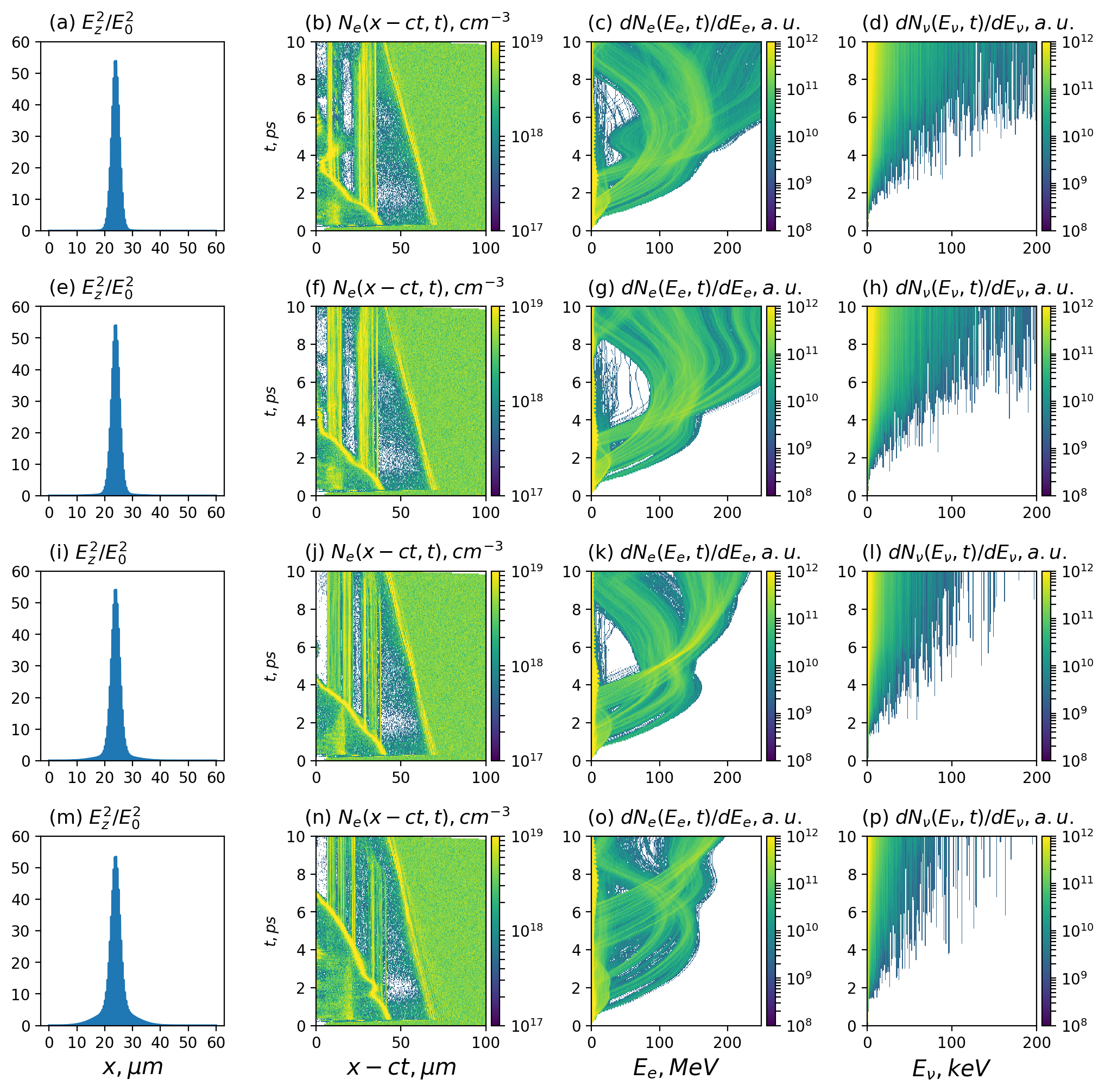}
\caption{\label{fig:pedestal}Comparison of simulation results for 11 fs laser pulses with 50 fs pedestals at different amplitudes. The given data is similar to Fig.~\ref{fig:50vs11}. 1st row --- pulse with $a_0=7.1$ and pedestal with $a_0=0.4$ (total energy 3.6 J), 2nd row --- pulse with $a_0=6.7$ and pedestal with $a_0=0.8$ (total energy 3.3 J), 3rd row --- pulse with $a_0=5.8$ and pedestal with $a_0=1.7$ (total energy 3.2 J), 4th row --- pulse with $a_0=4.2$ and pedestal with $a_0=3.3$ (total energy 4.6 J)}
\end{figure}

The simulation results are shown in Fig.~\ref{fig:pedestal}. It can be seen that a pedestal with an amplitude less than 10~\% of the amplitude of the main pulse does not noticeably change the acceleration dynamics, although it can cause some elongation of the accelerating cavity. The spectrum of electrons and generated photons change insignificantly. At larger amplitudes of the pedestal, however, the plasma wake generated by it interferes with the wake generated by the main pulse, which leads to a disruption of the acceleration process, a decrease in the number of high-energy electrons, and, consequently, a decrease in the efficiency of betatron generation.

\subsection{Influence of a prepulse}
In the next series of calculations, we studied the influence of a femtosecond scale prepulse on the acceleration process. The main pulse in these calculations had a Gaussian shape, a duration of 11 fs, and an amplitude $a_0 = 7.5$ (pulse energy 3.9 J). The prepulse also had a Gaussian shape and a duration of 5 fs. In the first series of calculations, its amplitude varied in the range from $a_0=0.8$ to $a_0=3.3$, and the arrival time was fixed and equal to 8.5 fs in front of the main pulse. Note that even for the most intense prepulse, its energy was only 0.34 J, i.e., about 10~\% of the energy of the main pulse.

\begin{figure}
\centering
\includegraphics[width=0.5\textwidth]{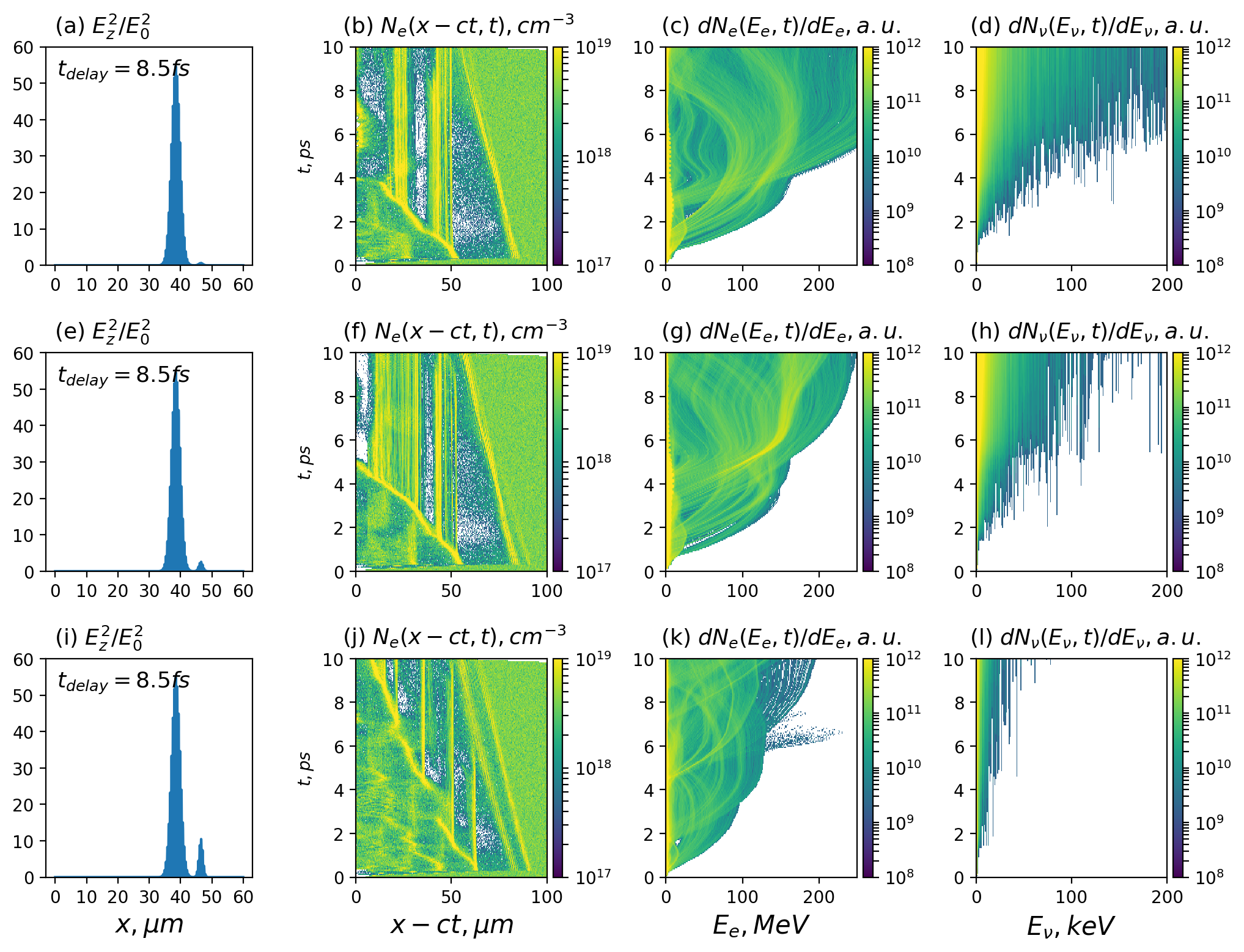}
\caption{\label{fig:prepulse_vs_a}Comparison of simulation results for 11 fs laser pulses with 5 fs prepulses with different amplitudes and located at 8.5 fs in front of the main pulse. The given data is similar to Fig.~\ref{fig:50vs11}. 1st row --- prepulse with $a_0=0.8$ (energy 0.02 J), 2nd row --- prepulse with $a_0=1.7$ (energy 0.08 J), and 3rd row --- prepulse with $a_0=3.3$ (energy 0.34 J)}
\end{figure}

The results of these calculations are shown in Fig.~\ref{fig:prepulse_vs_a}. From Fig.~\ref{fig:prepulse_vs_a} (a)-(d) it can be seen that a prepulse with a non-relativistic amplitude has little effect on the interaction process. The spectra of generated electrons and photons change insignificantly in its presence. In Fig.~\ref{fig:prepulse_vs_a} (e)-(h) for a prepulse with a slightly relativistic amplitude, an expansion of the accelerating cavity is observed, which leads to a deterioration in the acceleration process. The number of electrons with energies above 200 MeV decreases, and, as a result, the number of photons with energies above 100 keV decreases as well. Part of the spectrum of electrons and photons with lower energies, however, does not change so significantly. Finally, as seen at Fig.~\ref{fig:prepulse_vs_a} (i)-(l) more intense prepulse leads to a significant modification of the acceleration process caused by the interference of the plasma wake generated by them with the wake of the main pulse. The acceleration process in this case is strongly disturbed; the energy of the accelerated electrons is significantly reduced, as well as the energy of the generated photons.

In the second series of calculations, the amplitude of the prepulse was fixed at $a_0=3.3$, and its delay relative to the main pulse was varied in the range from 8.5 fs to 70 fs. Thus, we investigated whether it is possible to reduce the negative effect of a powerful prepulse by moving it away from the main pulse.

\begin{figure}
\centering
\includegraphics[width=0.5\textwidth]{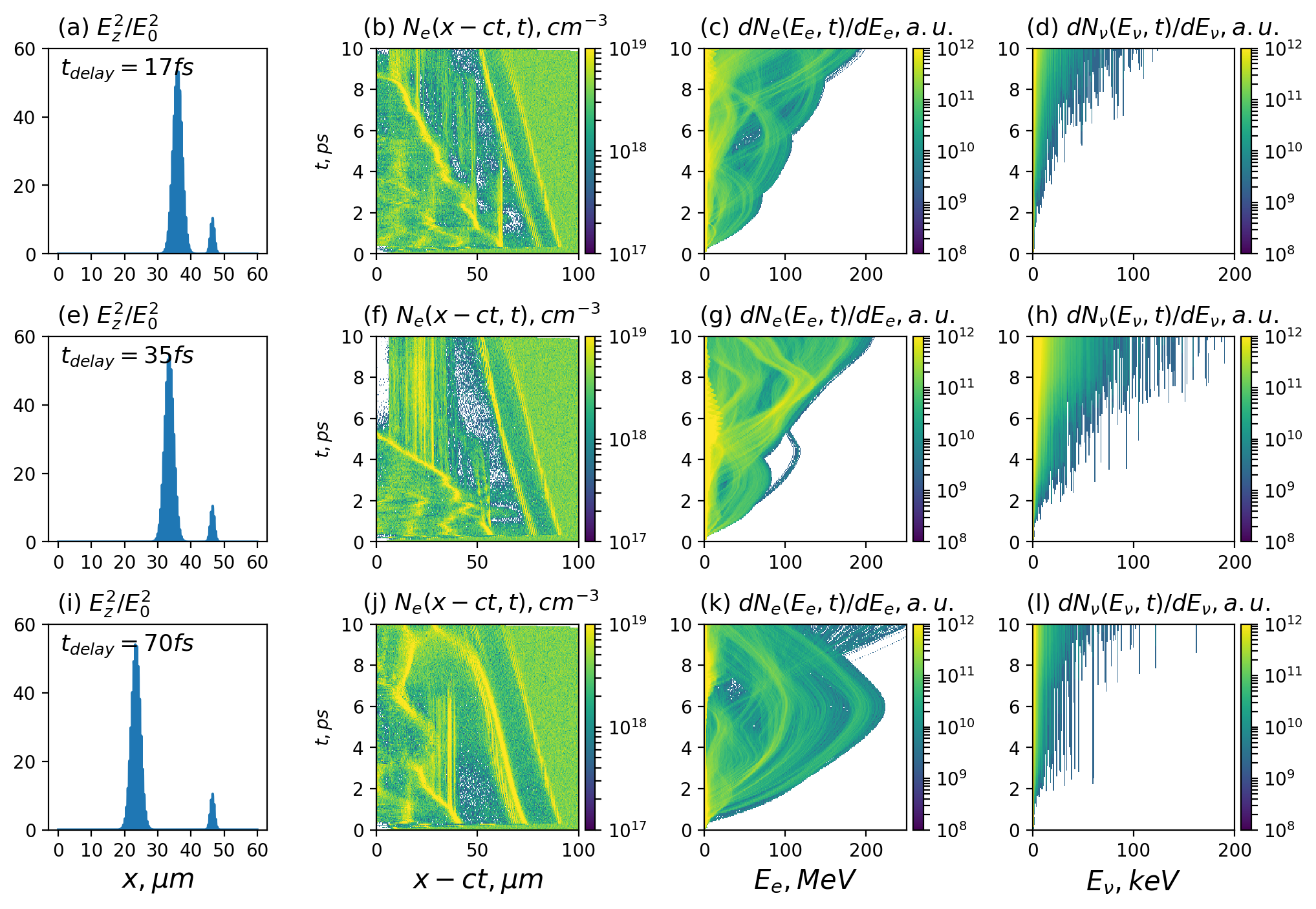}
\caption{\label{fig:prepulse_vs_delay}Comparison of simulation results for 11 fs laser pulses with 5 fs prepulses with fixed amplitude $a_0=3.3$ (energy 0.34 J) and variable delay $t_{\rm delay}$ in front of the main pulse. The given data is similar to Fig.~\ref{fig:50vs11}. 1st row --- prepulse with $t_{\rm delay}=17$ fs, 2nd row --- prepulse with $t_{\rm delay}=35$ fs, 3rd row --- prepulse with $t_{\rm delay}=70$ fs.}
\end{figure}

The results of this series of calculations are shown in Fig.~\ref{fig:prepulse_vs_delay}. It can be seen that the destructive influence of the interference of plasma wakes is observed in any case, including the one (Fig.~\ref{fig:prepulse_vs_delay} (i)-(l)) with a sufficiently large distance between the pulse and the prepulse, which exceeds the length of the plasma wave. The degree of this influence, however, varies nonmonotonically with the distance. The most destructive is the presence of a prepulse just before the main pulse.

\subsection{Influence of a postpulse}
In the next series of calculations, the effect of a postpulse on the acceleration process was studied. Just as in the case of the prepulse, the main pulse in these calculations had a Gaussian shape, a duration of 11 fs, and a constant amplitude $a_0 = 7.5$ (pulse energy 3.9 J). The postpulse had a Gaussian shape and a duration of 5 fs. In the first series of calculations, its amplitude was varied in the range up to $a_0=3.3$, and its delay after the main pulse was fixed and equal to 8.5 fs. In this case (not shown), the postpulse didn't noticeably influence the acceleration dynamics.

\begin{figure}
\centering
\includegraphics[width=0.5\textwidth]{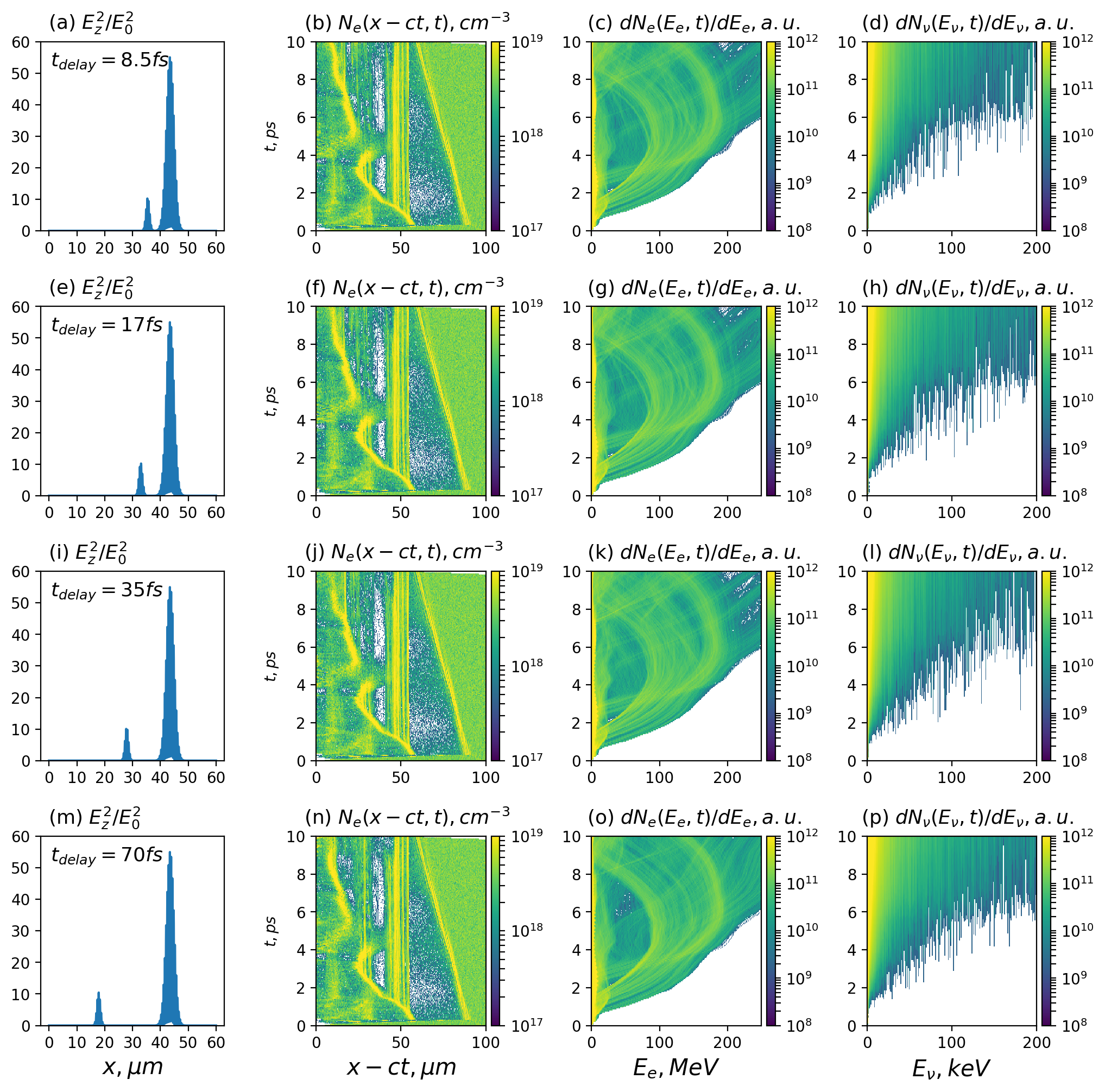}
\caption{\label{fig:postpulse_vs_delay}Comparison of simulation results for 11 fs laser pulses with 5 fs postpulses with fixed amplitude $a_0=3.3$ (energy 0.34 J) and variable delay $t_{\rm delay}$ after the main pulse. The given data is similar to Fig.~\ref{fig:50vs11}. 1st row --- postpulse with $t_{\rm delay}=8.5$ fs, 2nd row --- postpulse with $t_{\rm delay}=17$ fs, 3rd row --- postpulse with $t_{\rm delay}=35$ fs, 4th row --- postpulse with $t_{\rm delay}=70$ fs}
\end{figure}

In the second series of calculations, the amplitude of the prepulse was fixed at $a_0=3.3$, and its distance from the main pulse was varied in the range from 8.5 to 70 fs. The results of this series of calculations are shown in Fig.~\ref{fig:postpulse_vs_delay}. It can be seen that even in the case when the postpulse overlaps in space with the accelerated electron beam, the changes in the acceleration dynamics are insignificant, and the spectra of electrons and photons remain nearly the same.

\begin{figure*}
\centering
\includegraphics[width=1\textwidth]{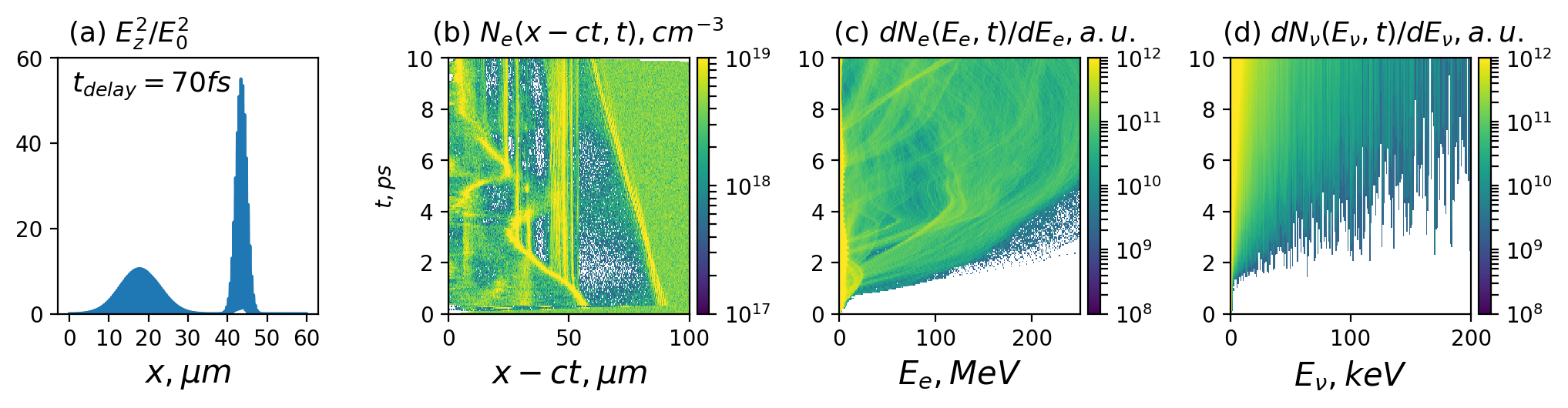}
\caption{\label{fig:longpostpulse}The result of simulation for a laser pulse with a duration of 11 fs with a postpulse with a duration of 40 fs, amplitude $a_0=3.3$ (energy 2.7 J), following the main pulse with a delay of 70 fs. The given data is similar to Fig.~\ref{fig:50vs11}}
\end{figure*}

This, however, turns out to be valid only for sufficiently short postpulses. Fig.~\ref{fig:longpostpulse} shows the calculation results for the case of a postpulse with a duration of 40 fs and an amplitude $a_0=3.3$ (energy 2.7 J), following the main pulse with a delay of 70 fs. In this case, the postpulse fields noticeably increase electron energies by direct acceleration and enhance their oscillations, both of which lead to an increase in the brightness of the betatron radiation, similar to the idea proposed recently by Xi Zhang et al. \cite{Zhang2015}.

\section{\label{sec:conclusions}Conclusions}

In conclusion, our study shows that the femtosecond structure of a laser pulse can have a significant effect on its interaction with underdense plasmas. The presence of a pedestal and prepulses can lead to the excitation of plasma wakes interfering with the wake from the main pulse. In this case, the energy of the accelerated electrons can decrease and the spectrum of the radiation generated by them can narrow. However, to observe such effects, the pedestal amplitude must exceed 10~\% of the main pulse amplitude, and the amplitude of the prepulses must exceed relativistic amplitude, that is, their energy must be more than a few percent of the main pulse energy. Moreover, in experiments the focusing properties of the various components of the pulse also play an important role. Due to the dependence of the nonlinear phase acquiring during the self-phase modulation on the local amplitude of the wave, the structure of the phase front for the main pulse, prepulses, and pedestal can differ significantly, which will lead to an effective decrease in the amplitude of the prepulses and the pedestal at the focus \cite{Khazanov2019, Martyanov2022}.

It also follows from our calculations that the presence of femtosecond postpulses, as a rule, does not significantly change the acceleration dynamics and the spectrum of the generated radiation. Moreover, it can have a positive effect for sufficiently long and intense postpulses located at an optimal distance after the main pulse. In that case they overlap with the accelerated electrons, which intensify their oscillations and, as a consequence, lead to an increase in the brightness of betatron radiation.

Thus, taking into account the femtosecond structure of ultrashort laser pulses obtained by compression after nonlinear self-phase modulation in thin plates can be important in interpreting the results of their interaction with matter, in particular, in the problem of generating betatron radiation by laser wakefield accelerated electrons.

\section{Acknowledgements}

This work was supported by the Ministry of Science and Higher Education of the Russian Federation (Agreement No. 075-15-2021-1361). The simulations were performed on resources provided by the Joint Supercomputer Center of the Russian Academy of Sciences.

\nocite{*}
\bibliography{betatron}

\end{document}